\documentclass[aps,pre,twocolumn,groupedaddress]{revtex4}

\usepackage{graphicx}

\begin{document}


\title{Linear scaling calculation of a $n$-type GaAs quantum dot}


\author{Shintaro Nomura}
\email[]{snomura@sakura.cc.tsukuba.ac.jp}
\affiliation{Institute of Physics, University of Tsukuba,
1-1-1 Tennodai, Tsukuba, Japan}
\author{Toshiaki Iitaka}
\email[]{tiitaka@riken.jp}
\affiliation{RIKEN (The Institute of Physical and Chemical Research),
2-1 Hirosawa, Wako, Japan}


\date{\today}

\begin{abstract}
A linear scale method for calculating electronic
properties of large and complex systems is introduced within a local density
approximation.
The method is based on the Chebyshev polynomial expansion
and the time-dependent method, which is tested in calculating
the electronic structure of a model $n$-type GaAs quantum dot.
\end{abstract}


\maketitle



\section{INTRODUCTION}

Linear scale methods for calculating the electronic structures have been
actively investigated in the last decade
because of increasing demands for predicting properties of
large and complex systems with computational cost linear scale
with respect to the system size $N$ \cite{Goedecker}.
There are several approaches for achieving linear scaling, such as
the divide-and-conquer (DC) method \cite{Yang91a},
the density-matrix minimization (DMM) method \cite{Li93},
the orbital minimization (OM) method \cite{Mauri93},
and the Chebyshev polynomial expansion (CPE) method \cite{Kosloff86,Silver96,Goedecker}.
Computational efficiency and applicability for specific systems have been
mostly tested based on the tight-binding (TB) formalism.
The DC method divides a system into subsystems in physical space
and obtains the density matrix for each subsystem.
This method is highly efficient if small localization region can be
chosen as subsystems, but this depends on the problems and becomes more
difficult for calculations based on the finite-difference (FD) formalism
with a large basis set. 
While the TB method is very successful in quantum chemistry, 
care must be taken for constructing appropriate basis set for a particular
problem \cite{Hoshi06}.  
A calculation based on the FD formalism ~\cite{Beck00} is straightforward
and is widely used for electronic structure calculations of semiconductors
and biochemical systems.
The DMM and OM methods, which require to store the whole density matrix and 
the whole Wannier functions, respectively,
suffer from their large memory requirements.
In the CPE method, the memory requirements are significantly reduced
because only small number of column vectors is required to be stored. 
Since neither division into subsystems or
the initial guess of the initial state is required,
the CPE method is straightforwardly applied to a wide variety of systems.
The other important advantage of the CPE method is
suitability for parallel implementation.
Because the most time-consuming part of the calculation is 
matrix-times-vector multiplication, where each column of the Hamiltonian
matrix can be treated as independent, communications between clusters
are minimized.
The CPE method is thus suitable for achieving linear scaling based
on the FD formalism with a large basis set.

In the CPE method the electron density is evaluated
by using a matrix representation of the Fermi-operator,
which is expanded in the Chebyshev matrix polynomials.
The so-called Gibbs oscillation in the zero temperature case
is suppressed by using finite-temperature Fermi operator.~\cite{Baer97,Baer98} 
In the tight binding approach, the linear scaling is obtained by
a truncated Hamiltonian which retains only matrix elements inside a
localization region ~\cite{Voter96}.
Reasonably small localization can be defined
for a tight-binding approach with, for example, atom-centered basis functions.
In the FD formalism, it is not obvious how to define a localization region
where basis functions are retained. Moreover, because the number of
basis functions within a localization region becomes much larger
than the tight binding approach,
the crossover point where the linear scaling approach is faster than
a conventional approach such as a conjugate gradient method
(CGM) becomes significantly larger. 

This leads us to utilize the other approach of calculating
the trace of a large matrix by using random vectors ~\cite{Hams00,Iitaka04}.
In calculating physical quantities such as energy, electron density,
or linear response function, the trace of a relevant operator $A$ needs to be 
calculated. If $A$ is expressed in terms of a basis set $\phi _{q},
q=1,..., N_{d}$ as $\rm tr[\it A] = \sum_{q=1}^{N_{d}} A_{qq}$,
the calculation of this part costs $O(N_{d}^{2})$
if the matrix is expanded in the Chebyshev matrix polynomials \cite{Goedecker}.
By introducing a random phase vector as defined by
$|\Phi \rangle \equiv \sum_{q=1}^N |q \rangle \xi_q $,
where $\{ |q \rangle \}$ is a basis set and
$\xi_q$ are a set of random phase variables,
the trace is evaluated at the cost of $O(N_{d})$ as given by
$\rm tr[\it A] = \langle\langle \Phi|A|\Phi \rangle\rangle$,
where 
$\langle{\langle{\cdot}\rangle}\rangle$ stands for statistical average.
The overall linear scaling is obtained by this method.
The random phase vector was shown to give results
with the smallest statistical error ~\cite{Iitaka04}.
This approach is also known to show a useful feature called
the self-averaging effect that the fluctuation in some physical quantities
decreases with increase in $N_{d}$ for sparse or banded matrices $A_{nm}$.
With a combination of this approach 
and the time-dependent method \cite{Iitaka94} (CPE-TDM),
linear response functions or electron density of states (DOS)
are calculated by integrating the time-dependent Schr\"{o}dinger equation 
without calculating eigenenergies or eigenstates.
The computational time of CPE-TDM scales as $O(N)$,
as compared with that of the conventional method such as
conjugate gradient method (CGM), which grows as $O(N^{2})$.
Thus CPE-TDM enables us to calculate electronic properties of large systems
which require prohibitively large computational time by CGM.
CPE-TDM was applied to calculate the optical properties of hydrogenated Si
nanocrystals containing atoms more than 10,000
within the empirical pseudopotential formalism \cite{Nomura97,Iitaka97},
the optical properties of carbon nanocrystals \cite{Kurokawa00a}
and polysilane \cite{Kurokawa00c},
and
the electron spin resonance spectrum of $s=1/2$ antiferromagnet Cu benzoate \cite{Iitaka03},
which have proved the advantages of CPE-TDM.
However, CPE-TDM has not been applied to calculation of
the electronic structure within a local density approximation (LDA).
Applications of a  linear scaling method with the self-consistent-field
level of theory are still very limited, but this level of calculation
using Gaussian basis sets
has been demonstrated to be practical. \cite{Weber04} 

In this paper, we report on an implementation of CPE-TDM for a large scale
calculation of the electronic structure of $n$-type GaAs quantum dot (QD)
\cite{Hawrylak93,Stopa96}
within a LDA based on a FD formalism and compare the results with a CGM.\\

\section{METHOD}

The model structure is a 20 nm-wide GaAs quantum well sandwiched by
undoped $\rm Al_{\it x}Ga_{1-\it x}As$ $(x=0.3)$ barriers, which 
confine the electrons with the effective-mass $m^{*}$ in the $z$ direction.
For QDs, the electrons are assumed to
be laterally confined to a harmonic oscillator with frequency $\omega_0$,
which may be created by a surface gate structure \cite{Stopa96} in experiments.
The electrons are assumed to be supplied from 5 nm-thick
Si-doped $\rm Al_{x}Ga_{1-x}As$ layer, located 20 nm above the GaAs
quantum well layer.
The Fermi-energy ($E_{F}$) is taken as the origin of the energy.
The Fermi-level pinning model is assumed \cite{Hirayama98}.
The number of the electrons in a QD is not fixed to an integer number and
is determined by $E_{F}$ and the potential energy.

The model Hamiltonian of the system within the LDA is
\begin{equation}
H ={{\bf p}^{\rm 2} \over 2\it m^*}+{1 \over 2} m^{*}\omega_0^{2}
({\it x}^{2}+{\it y}^{2})
+{\it V}_{c}\left({\it z}\right)+{\it V}_{H}\left({\bf r}\right)
+{\it V}_{x}\left({\bf r}\right)
\end{equation}
where $V_{c}(\it z\rm)$, $V_{H}(\bf r\rm)$, and $V_{x}(\bf r\rm)$
are the vertical confining potential, the Hartree potential,
and the exchange potential, respectively.
A 3D mesh of $64 \times 64\times 8$
is used for the calculation of the electron density,
and  $64\times 64\times 16$
is used for the calculation of the potentials.
The axis perpendicular to the quantum well layer is taken to be $z$-direction 
and the grid-spacing $\Delta x$ is fixed to be 5 nm.
The Hamiltonian is discretized in real-space by
the higher-order finite difference method \cite{Chelikowsky94A,Nomura04}.

The electron density at finite temperature is given by \cite{Kumar90}
\begin{equation}
n(\bf{r}) = \it \sum_{j} \phi_{j}^{*}(\bf{r}) \phi_{\it j}(\bf{r})
    \rm \it f((E_{j}-E_{F}))
\end{equation}
where $\phi_{j}$ and $E_{j}$ are the one-particle wave function and the
energy of the $j$th electron state, respectively, which are obtained
by CGM.
$f(E_{j}-E_{F}) ={1 \over {{e}^{\beta (E_{j}-{E}_{F})}+1}}$
is the Fermi distribution function at inverse temperature $\beta$. 
We use $\beta = 4000$ $\rm eV^{-1}$ corresponding to
the temperature $T= 2.9$ K.
The electron states above $E_{F}$ are partially occupied
due to this finite temperature effect.  
The introduction of finite temperature accelerate
convergence of the self-consistent-field loop.

In CPE-TDM, a random phase vector as defined by
$|\Phi \rangle \equiv \sum_{q=1}^N |q \rangle \xi_q $, where
$\xi_q$ are a set of random phase variables $\xi_q={e}^{i\phi_{q}}$,
is used as an initial state.
Here $\Phi$ is a $N_{x} \times N_{y} \times N_{z}$ column vector
for a system defined by a real-space uniform grid of
$N_{x} \times N_{y} \times N_{z}$.
The electron density $n(\bf r\rm)$ is extracted by the Fermi operator function
$f(H) ={1 \over {{e}^{\beta (H -{E}_{F})}+1}}$ as
\begin{equation}
n(\bf r\rm)=\langle \langle |\langle \Phi |\it f(H)|\bf r\rm \rangle {|}^{2}\rangle \rangle,
\end{equation}
where $\beta$ is connected to a real temperature. The Fermi operator
is evaluated by the Chebyshev polynomial expansion,
\begin{equation}
f(H)|\Phi\rangle=\sum_{k}\it a_{k}(\beta)T_{k}(H)|\rm\Phi\rangle.
\end{equation}

The length of the Chebyshev expansion for precision $10^{-D}$
is given by~\cite{Baer97, Baer98} 
\begin{equation}
P = {2 \over 3} (D-1) \beta \cdot \Delta E
\end{equation}
where $\Delta E = (E_{max} - E_{min})/2$.
We use $D=6$, $\Delta E=1.0$ eV, $\beta = 4000$ $\rm eV^{-1}$, giving $P=13333$. 
A calculation was also performed with $D=9$, and we find that the differences
in the total number of electrons ($N_{e}$) and the Hartree potential ($V_{H}$)
between the two cases of $D=6$ and $D=9$
were less than $1 \times 10^{-3}$ and $1 \times 10^{-6}$ eV, respectively.

The electron density is calculated with $j_{\rm max \it}$ sets of
$|\Phi\rangle$
as
$n(\bf r\rm)=\it \sum_{j=\rm 1}^{\it j=j_{\rm max \it}}
\left\langle \bf r\rm|\it f(H)|\rm\Phi_{\it j}\right\rangle/j_{\rm max \it}$.
The fluctuation for the random phase vector is ~\cite{Iitaka04, Baer04}
\begin{equation}
\delta H/L \approx {{\hbar \over{2 m^{*}(\Delta x)^{3}}}
 {\sqrt{2} \over{\sqrt{j_{max}N}}}},
\end{equation}
where $L=N_{x} (N_{y}) h$
as the number of meshes $N \rightarrow \infty$.
The statistical error decreases as $1/\sqrt{j_{max}}$ in general.

While it is known that other representation of a smoothed step function
such as a complementary error function yields improvements of degree
of polynomial expansion ~\cite{Liang03}, we use the Fermi operator
because this is physically correct for electronic structure calculations
at finite temperature. 
The Hartree and exchange potentials are calculated by using Eq. (3).
Therefore, it is not necessary to obtain eigenvalues or eigenfunctions.
The new solution of the potential $V_{\rm H}^{\rm new}(\bf r\rm)$
is combined with the solution obtained for the previous iteration by
$V_{\rm H}(\bf r\rm) =(1-\alpha)V_{\rm H}^{\rm old}(\bf r\rm)+
\it\alpha V_{\rm H\it}^{\rm new}(\bf r\rm)$.
Similarly, in order to reduce the statistical fluctuation,
$n(\bf r\rm)$ is combined with the density obtained
for the previous iteration by
$n(\bf r\rm) =(1-\gamma)\it n^{\rm old}(\bf r\rm)+
\it\gamma n^{\rm new}(\bf r\rm)$.
The parameter $\alpha$ is fixed to be 0.08, and
the parameter $\gamma$ is varied between 0.3 and 0.1.

A real-time Green's function $G({\omega}_{\ell}+i\eta)$ is calculated by
a time evolution method by solving a homogeneous Schr\"{o}dinger equation
numerically with an initial condition $\phi (q, t=0)=| q \rangle$
as~\cite{Iitaka97}
\begin{equation}
{\tilde{\phi}}_{\ell}(q, T) =(-i)\int_{0}^{T}dt' \phi (q, t'){e}^{i (\omega_{\ell}+i\eta)t'}
\end{equation}
\begin{equation}
\approx {1 \over {\omega }_{\ell}+i\eta -H} | q \rangle
\end{equation}
\begin{equation}
= G({\omega}_{\ell }+i\eta)|q \rangle .
\end{equation}

This method is as efficient as the CPE method with a carefully chosen
Gibbs damping factor.~\cite{Silver96,Baer97}
The DOS is then calculated at the cost of $O(N_{d})$ as given by

\begin{equation}
\rho (\omega)=-{1 \over \pi} \rm Im (Tr [\it G(\omega + i\eta )]),
\end{equation}
\begin{equation}
 =-{1 \over \pi} \rm Im \it (\sum_{q, q'}^{} \langle\langle{{e}^{i({{\phi}_{q} -{\phi}_{q'}})}}\rangle\rangle \langle{q | G(\omega + i\eta ) | q'}\rangle)
\end{equation}
\begin{equation}
=-{1 \over \pi} \rm Im\it(\langle \langle \ \left\langle{
\rm \Phi |\it G(\omega + i\eta )|\rm \Phi }\right\rangle\ \rangle \rangle).
\end{equation}

The DOS is calculated with $k_{\rm max \it}$ sets of $|\Phi\rangle$.
The energy resolution $\eta$ is chosen to be 0.25 meV.
It should be noted that $k_{\rm max \it}$ used for
calculating the DOS can be independently chosen
from $j_{\rm max \it}$ for each self-consistent iteration procedures.\\

\begin{figure}
\includegraphics[width=7.5cm]{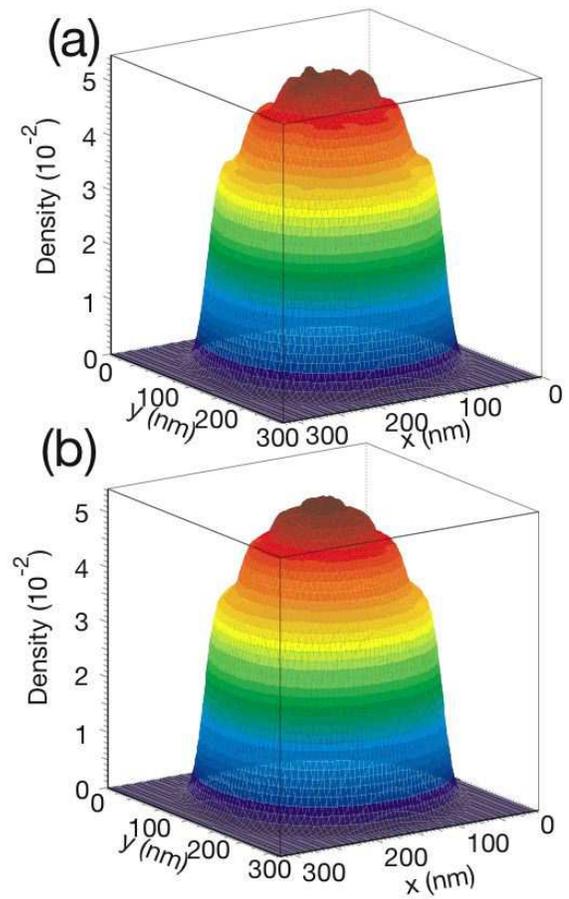}
\caption{\label{dens}
(a) The electron density distribution obtained by CPE-TDM.
128 sets of random vectors are used at each self-consistent
iteration procedure.
(b) The electron density distribution obtained by CGM.
}
\end{figure}

\section{RESULTS AND DISCUSSIONS}

Model calculations are performed for GaAs QDs containing about 77
electrons. We take $\omega_0=3$ meV for a typical GaAs QD \cite{Hawrylak93}.
The number of the self-consistent iterations is fixed to 100
for both the CGM and CPE-TDM calculations.
The potential is converged to
$|V_{\rm H}(\bf r\it) - V_{\rm H}^{\rm new}(\bf r\rm)| < 0.003$ meV for the CGM
calculation.
The electron density distributions are shown in Fig.~\ref{dens} for CPE-TDM
with $j_{\rm max}=128$ and CGM.
The calculated electron density distribution reasonably agrees
with the result by a CGM within the statistical fluctuations.
The Friedel-type spatial oscillations of the electron density \cite{Luscombe92}
are reproduced in both the results by the CPE-TDM and CGM.

\begin{figure}
\includegraphics[width=7.5cm]{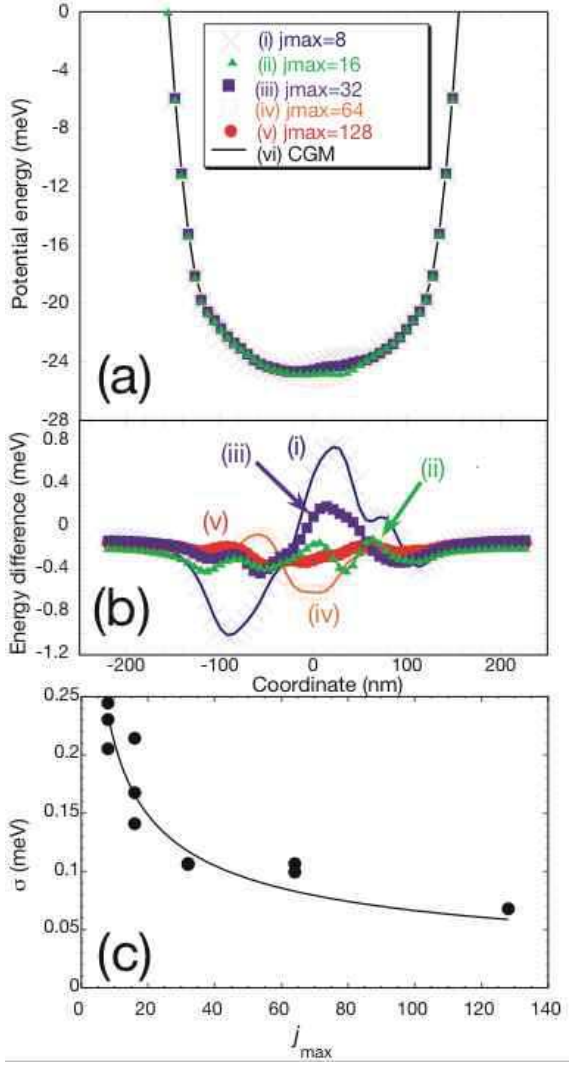}
\caption{\label{pot}
(a) Cross-sectional views of the calculated Hartree potentials 
on the plane at the center of the quantum well layer obtained
by CPE-TDM ($V_{\rm H}^{\rm CPE-TDM}(\bf r\it)$)
with (i) 8, (ii) 16, (iii) 32, (iv) 64, and (v) 128 sets of
random phase vectors for extracting $n(\bf r\rm)$
at each self consistent iteration procedure,
and (vi) $V_{H} (\bf x\rm)$ obtained by CGM.
(b) Differences of obtained Hartree potentials
$V_{\rm H}^{\rm CPE-TDM}(\bf r\it) - V_{\rm H}^{\rm CGM}(\bf r\rm)$ 
with (i) 8, (ii) 16, (iii) 32, (iv) 64, and (v) 128 sets of
random phase vectors.
(c) Standard deviations of the calculated Hartree potentials 
depending on the number of random phase vectors at each self consistent
iteration procedure $j_{\rm max}$.
The best fitted curve proportional to $1/\sqrt{j_{\rm max}}$ is also shown.
}
\end{figure}

The calculated Hartree potentials reasonably agree with the potential
obtained by the CGM as shown in Fig.~\ref{pot} (a). 
Differences of the calculated Hartree potentials with that by the CGM
are examined in Fig.~\ref{pot} (b). The absolute values of the difference
are smaller than 1.0 and 0.4 meV for $j_{\rm max}=8$ and 16, respectively. 
Figure~\ref{pot} (c) shows that the standard deviations of the differences
of the calculated Hartree potentials follows the curve proportional
to  $1/\sqrt{j_{\rm max}}$ as expected.

\begin{figure}
\includegraphics[width=7cm]{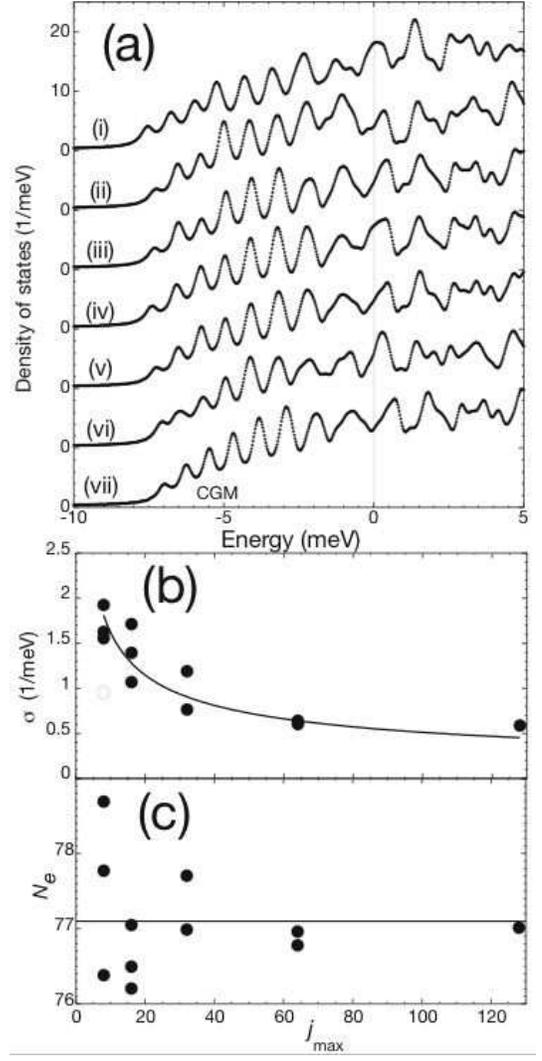}
\caption{\label{dos}
(a) Density of states ($\rho (\omega)$) obtained by
CPE-TDM with $j_{\rm max \it}$ = $k_{\rm max \it}$ = (i) 8, (ii) 16, (iii) 32, (iv) 64, and (v) 128 sets of
random phase vectors for extracting $n(\bf r\rm)$
at each self consistent iteration procedure and
for evaluating $\rho (\omega)$. (vi) $\rho (\omega)$
obtained by CPE-TDM with 8 sets of random phase vectors
for extracting $n(\bf r\rm)$ and
$k_{\rm max \it}$ = 64 sets of random phase vectors
for evaluating $\rho (\omega)$. (vii) $\rho (\omega)$ obtained by CGM.
(b) Standard deviations of the difference of the peak heights of the DOS
obtained by CPE-TDM and by CGM depending on $j_{\rm max}$ (solid circles).
Standard deviation for CPE-TDM with $j_{\rm max}=$ 8 at each self consistent
iteration procedure and $k_{\rm max \it}$=64 for evaluating $\rho (\omega)$
is shown (open circle)
The best fitted curve proportional to $1/\sqrt{j_{\rm max}}$ is also shown.
(c) Total number of electrons ($N_{e}$) depending on $j_{max}$. 
The horizontal line shows $N_{e} = 77.1$ obtained by CGM.
}
\end{figure}

The calculated DOS are shown in Fig.~\ref{dos}.
For CPE-TDM, the self-consistent iteration procedures are performed with
$j_{\rm max \it} = 8$, 16, 32, 64, and 128. The same number of
random phase vectors are used
for evaluating $\rho (\omega)$ except for the case of
$j_{\rm max}=8$ where $\rho (\omega)$ is evaluated with
$k_{\rm max}=8$ and 64.
It can be seen that the statistical fluctuations decrease with increase
in $j_{\rm max \it}$ in calculating $\rho (\omega)$.
There are two types of the fluctuations observed in Fig.~\ref{dos}.
One is the fluctuation in the peak energy positions,
and the other is the fluctuation in the peak heights.
The former can be reduced by increasing $j_{\rm max \it}$
and by decreasing the mixing parameter $\gamma$.
The latter also depends on $k_{\rm max \it}$.
In fact, the fluctuations in the peak heights are reduced by increasing
$k_{\rm max \it}$ from 8 to 64 with small changes
in the peak energy positions in the case of $j_{\rm max \it} = 8$
as shown in Fig.~\ref{dos} (b).
Figure~\ref{dos} (b) shows that the standard deviations of the peak heights
also follows the curve proportional to $1/\sqrt{k_{\rm max}}$.

Finally we note that the statistical fluctuation of
the total number of electrons ($N_{e}$) is
smaller than that of DOS because of the self-averaging effect.
Figure~\ref{dos} (c) shows calculated $N_{e}$ depending on $j_{\rm max \it}$. 
The statistical errors as compared with $N_{e} = 77.1$ by CGM
are as small as 2\% for $j_{\rm max \it}=8$,
which indicates that the self-averaging effect is effective for
a sparse banded matrix case as illustrated in this paper.

Our linear scale method opens up possibilities for calculating
the electronic and optical properties of large and complex systems, 
such as QD arrays with interaction between QDs and devices employing the Rashba type spin-orbit interaction \cite{Koga04}.
It should also be possible to calculate the electronic structure of
nanostructures within a LDA with \it ab initio \rm pseudopotentials.
Because the Green's function can be effectively estimated by CPE-TDM,
the properties of the electronic system such as the DC and Hall conductivities,
and the optical absorption spectra, are obtained within $O(N)$
computational costs.

In conclusions, it has been demonstrated that CPE-TDM can be applied to
a large scale calculation of a model QD within a LDA based on a FD formalism
despite the presence of the statistical fluctuations of
the calculated quantities originated from the random phase vectors.

\bibliography{cheby,Magbib}

\end{document}